\documentclass[prb,letterpaper,twocolumn,aps,floatfix]{revtex4-1}
\usepackage{graphicx}
\usepackage{xcolor}
\usepackage{amsmath}
\usepackage{amsfonts}
\usepackage[small,bf]{subfigure}
\newcommand{\beq}{\begin{equation}}
\newcommand{\eeq}{\end{equation}}
\newcommand{\bk}{{{\bf{k}}}}

\newcommand{\cK}{{{\cal K}}}

\newcommand{\br}{{{\bf{r}}}}

\newcommand{\bG}{{{\bf{G}}}}

\newcommand{\bq}{{\bf{q}}}

\newcommand{\beqa}{\begin{eqnarray}}
\newcommand{\eeqa}{\end{eqnarray}}

\newcommand{\btau}{{\boldsymbol \tau}}

\newcommand{\ra}{\rightarrow}

\newcommand{\cD}{{\cal D}}
\newcommand{\cL}{{\cal L}}
\begin{document}
\title{Absence of localization in Weyl semimetals}
\author{Jinmin Yi}
\affiliation{Department of Physics and Astronomy, University of Waterloo, Waterloo, Ontario 
N2L 3G1, Canada} 
\affiliation{Perimeter Institute for Theoretical Physics, Waterloo, Ontario N2L 2Y5, Canada}
\author{A.A. Burkov}
\affiliation{Department of Physics and Astronomy, University of Waterloo, Waterloo, Ontario 
N2L 3G1, Canada} 
\affiliation{Perimeter Institute for Theoretical Physics, Waterloo, Ontario N2L 2Y5, Canada}
\date{\today}
\begin{abstract}
One of the fundamental facts of condensed matter physics is that sufficient amount of disorder always turns a Fermi liquid metal into an Anderson insulator: a compressible, but non-conducting phase of matter. Recently, topological semimetals have emerged as another way a metallic phase may be realized. In this paper we point out that, unlike ordinary metals, at least some topological semimetals are immune to localization, provided certain conditions are satisfied. We present several physical arguments, based on diagrammatic perturbation theory and Keldysh field theory, as well as decorated domain wall construction, to back up this claim. 
\end{abstract}
\maketitle
\section{Introduction}
\label{sec:1}
Recent years have witnessed an expansion of topological concepts into the physics of metals.~\cite{Volovik03,Volovik07,Haldane04,Weyl_RMP}
This has in part been driven by the discovery of topological semimetals,~\cite{Murakami07,Wan11,Burkov11-1,Burkov11-2} which may be viewed as a new way 
a metallic phase can arise. 
An ordinary Fermi liquid metal is the result of non-integer number of electrons per unit cell per spin (filling), which translates into incompletely filled 
bands and thus a Fermi surface of gapless excitations, which encloses a volume in momentum space, directly related to the fractional part of the filling. 
This statement, known as Luttinger's theorem, has recently been understood to have topological origin.~\cite{Oshikawa00,Hastings04,Furuya17,Cheng16,Cho17,Metlitski18,Jian18,Song21,Else21,Gioia21,Wang21}

Topological semimetals, on the other hand, arise at integer filling, which normally corresponds to an insulator, as intermediate phases between insulators of different 
electronic structure topology.~\cite{Murakami07,Burkov11-1,Burkov11-2}
It is possible to understand topological properties of both ordinary metals and topological semimetals using the concept of unquantized anomalies.~\cite{Gioia21,Wang21,Sau24,Hughes24}
These are topological terms, but with non-integer and continuously-tunable coefficients, which encode long-range quantum entanglement and make the presence of gapless excitations or topological order~\cite{Wang20,Kane22,Yi23} (in the absence of broken symmetries) necessary. 

Impurity scattering is an essential and inevitable part of the physics of metals. Its effects range from relatively simple classical phenomena, e.g. the finite conductivity of a metal at low temperatures is exclusively the result of impurity scattering, to highly nontrivial effects, involving quantum interference of electron waves in the presence of a random potential.~\cite{Lee85,Altland10,Kamenev_book}
Of particular interest is the interplay of impurity scattering effects with topology. This interplay, for example, is behind the physics of the integer quantum Hall 
effect, where disorder-induced Anderson localization, combined with nontrivial Landau level topology, leads to the spectacularly precise quantization of the Hall 
conductance. 

In this paper we explore the effects of strong disorder in topological semimetals, focusing on the simplest (theoretically), yet in many ways the most interesting representative, the magnetic Weyl semimetal.~\cite{Burkov11-1,Burkov11-2}
This problem has been addressed before by a number of authors.~\cite{Arovas13,Brouwer14,Syzranov15,Wang15,Altland15,Pixley15,Xie15,Altland16,Hughes16,Pixley16,Brouwer16,Khalaf17,Syzranov18,Altland18,Mirlin19,Behrends24,Alisultanov24}
While there is some controversy regarding what happens at weak disorder, it is generally believed that at least 
at strong enough disorder a finite density of states appears at the Fermi energy and the Weyl semimetal thus turns into a compressible metal with 
diffusive transport. 
An important question is the fate of this diffusive metal when the disorder strength is increased even further. 
Does it eventually become localized, as an ordinary three-dimensional (3D) metal would, or not? 

A single Weyl node is anomalous and may only exist on the surface of a four-dimensional (4D) 
quantum Hall insulator. This means that a single Weyl fermion can not be localized. This, in turn, implies that in a Weyl semimetal with pairs of nodes of opposite 
chirality localization may only arise from inter-node scattering. Naively, one would then be tempted to say that at strong enough inter-nodal scattering a diffusive 
Weyl metal turns into an Anderson insulator.

In this paper we will demonstrate that the reality is a bit different and more interesting than the above argument would suggest. 
Weyl semimetal is distinct from an ordinary 3D metal in that it can not be localized by any amount of disorder, provided certain conditions, detailed below, are satisfied. 
The necessity for extra conditions arises from the fact that the question ``Can Weyl semimetal be localized by strong disorder?" is not well-posed as it stands. 
Since the Weyl nodes are not pinned to any special points in the Brillouin zone (BZ), it is guaranteed that a sufficiently general and sufficiently strong impurity potential 
will move them. Strong enough disorder will then always annihilate the nodes, producing a localized Anderson insulator. 

In order to avoid the somewhat trivial answer above, we ask the question differently. 
Namely, it is well-known that, at weak disorder, a magnetic Weyl semimetal may be viewed as an intermediate phase between an ordinary 3D insulator and an integer quantum Hall insulator with the Hall conductance of $e^2/h$ per atomic plane. 
We may then ask what happens to this intermediate phase at strong disorder, as long as it is not so strong as to eliminate the integer quantum Hall insulator phase 
altogether. 
In this case, we find that the intermediate phase, which we may still call Weyl semimetal, even though the Weyl nodes themselves are always destroyed by strong enough 
disorder, is never localized. 
We back up this claim by two different arguments, namely an argument based on the matrix nonlinear sigma model (NLSM) description of a 
disordered Weyl semimetal,~\cite{Wang15,Altland15,Altland16} as well as a decorated domain wall construction.~\cite{Fu12,Ma22,Ma23}
We note that the same conclusion was reached in an earlier study of a related, but not identical, problem of the transition between an axion insulator and a trivial insulator in 3D.~\cite{Stern21}

The rest of the paper is organized as follows. 
In Section~\ref{sec:2} we present diagrammatic perturbation theory of a disordered Weyl semimetal, with scalar gaussian-distributed impurity potential, 
ignoring quantum interference effects. We demonstrate that even at strong disorder, which creates a finite density of states at the Fermi energy and destroys the nodes, the Hall conductivity remains unchanged and equal to that of the clean Weyl semimetal. 
In Section~\ref{sec:3} we use the results of Section~\ref{sec:2} to develop a matrix NLSM description of a strongly-disordered Weyl semimetal, which describes 
quantum interference phenomena, leading to localization in ordinary metals. We demonstrate that 
a topological term with an unquantized continuously-tunable coefficient in this NLSM prevents localization. 
In Section~\ref{sec:4} we present another argument against localization in a disordered Weyl semimetal, based on a decorated domain wall construction. 
We conclude in Section~\ref{sec:5} with a discussion of our results. 

\section{Diagrammatic perturbation theory of a disordered Weyl semimetal}
\label{sec:2}
To develop the NLSM field theory, it is useful to start from a more straightforward diagrammatic perturbation theory of a disordered Weyl semimetal.
This gives a detailed picture of noninteracting diffusion modes, whose interactions, possibly leading to localization, are then taken into account within the NLSM formalism.
We begin with a minimum microscopic model for a magnetic Weyl semimetal with a pair of nodes, which is given by~\cite{Burkov11-1,Burkov11-2}
\beq
\label{eq:5}
H = v_F (\tau^x k_x + \tau^y k_y) + m(k_z) \tau^z, 
\eeq
where
\beq
\label{eq:6}
m(k_z) = m_0[\cos(k_z) - \cos(\cK)], 
\eeq
$\tau^a$ are Pauli matrices and we will use $\hbar = c = e = 1$ units throughout. 
This exhibits a pair of Weyl nodes located on the $z$-axis in momentum space at $k_z = \pm \cK$. 
We have ignored high-energy states here, however their contribution is important when calculating the Hall conductivity,~\cite{Burkov14-2}
and will be explicitly included then. 

We will start by discussing non-topological low-energy physics of the disordered Weyl semimetal, which involves impurity scattering and
ordinary transport. For this, it is sufficient to consider each node separately and simplify Eq.~\eqref{eq:5} even further. 
Let us focus on the right-handed Weyl node for concreteness, whose Hamiltonian is given by
\beq
\label{eq:7}
H = v_F \btau \cdot \bk, 
\eeq
where we have taken the Fermi velocity to be isotropic for simplicity. 
The energy eigenvalues and eigenvectors of the right-handed Weyl Hamiltonian are given by
\beq
\label{eq:8}
\epsilon_s(\bk) = s v_F k \equiv s \epsilon_{\bk}, 
\eeq
where $s = \pm$ and 
\beq
\label{eq:9}
| z^s_\bk \rangle = \frac{1}{\sqrt{2}} \left(\sqrt{1 + \frac{s k_z}{k}}, s e^{i \varphi} \sqrt{1- \frac{s k_z}{k}} \right)^T, 
\eeq
where $e^{i \varphi} = \frac{k_x + i k_y}{\sqrt{k_x^2 + k_y^2}}$. 

We will assume a scalar disorder potential, which is independent of the pseudospin index and is gaussian-distributed
\beq
\label{eq:10}
\langle V(\br) V(\br') \rangle = \gamma^2 \delta(\br - \br'). 
\eeq
Physically, this corresponds to a potential, which varies slowly on the scale of a unit cell of a clean crystal. 
This is both physically justified and simplifies the calculations, but our main results are independent of this assumption. 

\begin{figure}[t]
\centering
\includegraphics[width=\linewidth]{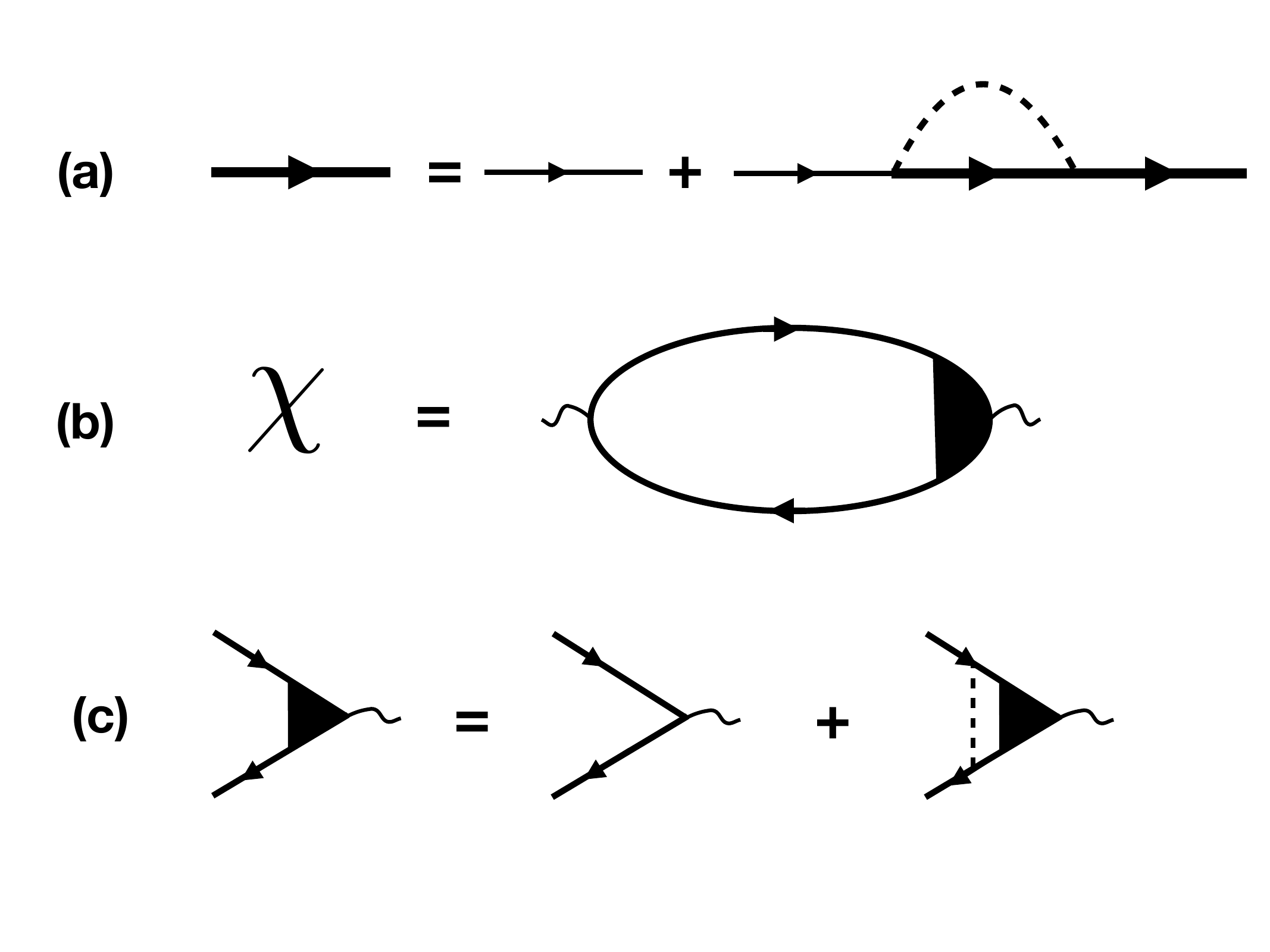}
\caption{Diagrammatic representation of (a) SCBA Green's function. Thin line represents the bare Green's function, thick line is the SCBA impurity-averaged Green's function and the dashed line represents the disorder potential correlator $\langle V(\br) V(\br') \rangle = \gamma^2 \delta(\br - \br')$. (b) Density response function $\chi$. 
(c) Equation for the diffusion vertex $\cD$.}
\label{fig:2}
\end{figure}

A reasonable description of the transport properties of any metal may be obtained by a combination of the self-consistent Born approximation (SCBA) 
for the electron self-energy due to impurity scattering and ladder vertex corrections for the density response bubble diagram, as shown in Fig.~\ref{fig:2}. 
This approximation respects all conservation laws, while ignoring quantum interference phenomena, which will be taken into account later using the NLSM formalism.

The retarded SCBA electron self-energy is given by
\beq
\label{eq:11}
\Sigma^R_s(\bk, \omega) = \gamma^2 \int \frac{d^3 k'}{(2 \pi)^3} \sum_{s'} |\langle z^s_\bk | z^{s'}_{\bk '} \rangle |^2 G^R_{s'}(\bk', \omega), 
\eeq
where 
\beq
\label{eq:12}
G^R_s(\bk, \omega) = \frac{1}{\omega - s \epsilon_\bk - \Sigma^R_s(\bk, \omega)}, 
\eeq
is the retarded disorder-averaged electron Green's function. We will assume an undoped stoichiometric Weyl semimetal henceforth, which is 
why the Fermi energy is set to zero in Eq.~\eqref{eq:12}. The doped Weyl semimetal case was discussed before in Ref.~\onlinecite{Burkov14-2}. 

If we look for a $\bk$- and $\omega$-independent solution of Eq.~\eqref{eq:11}, it is clear that $\Sigma^R_+ = \Sigma^R_-$. 
Then, the real part of the self-energy simply corresponds to an overall shift of the band dispersion and should be discarded if we keep the band filling 
constant. This means, in particular, that disorder potential does not shift the locations of the Weyl nodes. Note that this is generally only true for a scalar 
disorder potential. 

The imaginary part may be expressed in terms of the momentum relaxation time $\tau$ as
\beq
\label{eq:13}
\textrm{Im} \Sigma^R_+ = \textrm{Im} \Sigma^R_- = - \frac{1}{2 \tau}, 
\eeq
which gives the following self-consistent equation for the relaxation time
\beq
\label{eq:14}
1 = \gamma^2 \int_0^{k_0} \frac{d^3 k}{(2 \pi)^3} \frac{1}{v_F^2 k^2 + \frac{1}{4 \tau^2}},
\eeq
where $k_0$ is a cutoff momentum, which is of the order of the reciprocal lattice vector. 
This gives
\beq
\label{eq:15}
1 = \frac{\gamma^2 k_0}{2 \pi^2 v_F^2} \left[1 - \frac{1}{2 k_0 \ell} \textrm{arctan}(2 k_0 \ell) \right], 
\eeq
where $\ell = v_F \tau$ is the mean free path. 
This equation has a nontrivial solution once $\gamma$ exceeds a critical value, given by
\beq
\label{eq:16}
\gamma_c = \sqrt{\frac{2 \pi^2 v_F^2}{k_0}}.
\eeq
Solving Eq.~\eqref{eq:15} for the scattering rate in the vicinity of the transition for $\gamma > \gamma_c$ gives
\beq
\label{eq:16.1}
\frac{1}{\tau} \approx \frac{4 v_F k_0}{\pi} \left(1 - \frac{\gamma_c^2}{\gamma^2}\right). 
\eeq
Henceforth we will assume that $\gamma > \gamma_c$. This means that disorder has generated a finite density of states $g = 1/ 2\pi \gamma^2 \tau$ 
at the Fermi energy and the 
transport at time and length scales, longer than $\tau$ and $\ell$, is diffusive, as in an ordinary metal.~\cite{Syzranov15,Altland15,Altland16}
The existence of a sharp transition at $\gamma = \gamma_c$ has been disputed in Refs.~\onlinecite{Pixley15}, \onlinecite{Pixley16} and others, which pointed out that 
rare-region effects may give rise to a finite density of states even at an arbitrarily weak disorder. This claim was, however, challenged in Ref.~\onlinecite{Altland18}, which pointed out that Weyl nodes may still survive, even when rare-region effects are taken into account. 
We do not have anything new to say about this issue here and it is irrelevant for the question of localization, which is the main focus of our paper. 
What is important for us is just the existence of a diffusive transport regime for some interval of values of the disorder strength $\gamma$. 
We also note that, as seen from Eq.~\eqref{eq:16.1}, there exists an extended regime in which $1/\tau < v_F k_0$, i.e. the scattering rate is 
smaller than the bandwidth. This is the regime we are interested in. When $1/\tau > v_F k_0$, momentum-space description 
of transport becomes meaningless and all states may be expected to be localized in deep minima of the impurity potential. 
We are interested in the regime where it is still meaningful to speak about localization due to quantum interference. 

The full information about the electromagnetic response of our system is contained in the generalized density response function $\chi_{a b}(\bq, \Omega)$, 
where $a, b = 0, x, y, z$, corresponding to the electric charge and three components of the pseudospin. 
Since the current operator for a Weyl fermion is proportional to the pseudospin, $\chi_{a b}$ also contains information about all the components of the conductivity tensor. 
Neglecting quantum interference corrections, the response function is given by the bubble diagram with ladder vertex corrections, as shown in Fig.~\ref{fig:2}. 
It is useful to express it as a sum of two physically-distinct contributions
\beq
\label{eq:17}
\chi_{a b}(\bq, \Omega) = \chi^I_{a b}(\bq, \Omega) + \chi_{a b}^{II}(\bq, \Omega), 
\eeq
where 
\beq
\label{eq:18}
\chi_{a b}^I(\bq, \Omega) = \Omega \int_{-\infty}^{\infty} \frac{d \omega}{2 \pi i} \frac{d n_F(\omega)}{d \omega} P_{a b}(\bq, \omega - i \eta, \omega + \Omega + i \eta), 
\eeq
\beqa
\label{eq:19}
\chi_{a b}^{II}(\bq, \Omega)&=& \int_{-\infty}^{\infty} \frac{d \omega}{2 \pi i} n_F(\omega) \left[P_{a b}(\bq, \omega + i \eta, \omega + \Omega + i \eta) \right. \nonumber \\
&-& \left. P_{a b}(\bq, \omega - i \eta, \omega + \Omega - i \eta)\right].
\eeqa
Here $n_F(\omega)$ is the Fermi-Dirac distribution function and $\eta = 0+$. 
The $4 \times 4$ matrices $P$, which appear in the expressions above, are given by
\beq
\label{eq:20}
P(\bq, -i \eta, \Omega + i \eta) = \frac{1}{\gamma^2} I^{R A}(\bq, \Omega) \cD(\bq, \Omega), 
\eeq
and 
\beq
\label{eq:21}
P(\bq, \omega \pm i \eta, \omega + \Omega \pm i \eta) = I^{R R (A A)}(\bq, \omega, \Omega),
\eeq
where 
\beqa
\label{eq:22}
&&I^{R A}_{a b}(\bq, \Omega) = \frac{\gamma^2}{2} \tau^a_{\sigma_2 \sigma_1} \tau^b_{\sigma_3 \sigma_4} \nonumber \\
&\times&\int \frac{d^3 k}{(2 \pi)^3} G^R_{\sigma_1 \sigma_3}\left(\bk + \frac{\bq}{2}, \Omega\right) G^A_{\sigma_4 \sigma_2}\left(\bk - \frac{\bq}{2}, 0\right),
\eeqa
and 
\beqa
\label{eq:23}
&&I^{R R (A A)}_{a b}(\bq, \omega, \Omega) = \frac{1}{2} \tau^a_{\sigma_2 \sigma_1} \tau^b_{\sigma_3 \sigma_4} \nonumber \\
&\times&\int \frac{d^3 k}{(2 \pi)^3} G^{R (A)}_{\sigma_1 \sigma_3}\left(\bk + \frac{\bq}{2}, \omega + \Omega\right) 
G^{R (A)}_{\sigma_4 \sigma_2}\left(\bk - \frac{\bq}{2}, \omega \right), \nonumber \\
\eeqa
with summation over repeated spin indices made implicit. 
\beq
\label{eq:24}
\cD(\bq, \Omega) = [1 - I^{RA}(\bq,\Omega)]^{-1}, 
\eeq
is the diffusion propagator, or diffuson. 
The disorder-averaged SCBA Green's functions, which appear in Eqs.~\eqref{eq:22}, \eqref{eq:23}, may be expressed as
\beq
\label{eq:25}
G^{R (A)}_{\sigma_1 \sigma_2}(\bk, \omega) = G^{R (A)}_s(\bk, \omega) \delta_{\sigma_1 \sigma_2} + \bG^{R (A)}_t(\bk, \omega) \cdot \btau_{\sigma_1 \sigma_2}, 
\eeq
where 
\beqa
\label{eq:26}
G^{R (A)}_s(\bk, \omega)&=&\frac{1}{2} \left(\frac{1}{\omega - \epsilon_\bk \pm \frac{i}{2 \tau}} + \frac{1}{\omega + \epsilon_\bk \pm \frac{i}{2 \tau}} \right), \nonumber \\
\bG^{R (A)}_t(\bk, \omega)&=&\frac{\bk}{2 k} \left(\frac{1}{\omega - \epsilon_\bk \pm \frac{i}{2 \tau}} - \frac{1}{\omega + \epsilon_\bk \pm \frac{i}{2 \tau}} \right). \nonumber \\
\eeqa
Here the $s$ and $t$ subscripts denote singlet and triplet components of the Green's function respectively. 
Physically, $\chi^I(\bq, \Omega)$ describes dissipative response, involving multiple impurity scattering events at the Fermi energy. 
$I^{RA}$ corresponds to a single scattering event, while $\cD$ describes multiple scatterings, which lead to diffusion at long distances and long times. 
In contrast, $\chi^{II}(\bq, \Omega)$ corresponds to nondissipative response, involving all filled states below the Fermi energy and is largely independent 
of disorder, assuming $1/\tau$ is smaller than characteristic band energy scales.

After a straightforward calculation, we obtain 
\beq
\label{eq:27}
I^{RA}_{00}(\bq, \Omega) = \frac{i}{2 q \ell} \textrm{ln}\left(\frac{1 - i \Omega \tau - i q \ell}{1 - i \Omega \tau + i q \ell} \right). 
\eeq
Expanding this to leading nonvanishing order in $\Omega \tau$ and $q \ell$, which gives
\beq
\label{eq:28}
I^{RA}_{00}(\bq, \Omega) \approx 1 + i \Omega \tau - D \bq^2 \tau, 
\eeq
we obtain the expected result for the long-wavelength low-frequency electric charge density response function
\beq
\label{eq:29}
\chi_{00}(\bq, \Omega) = g \frac{D \bq^2}{D \bq^2 - i \Omega}, 
\eeq
where $g = 1/2 \pi \gamma^2 \tau$ is the static compressibilty, which is nonzero due to strong disorder, unlike in a clean Weyl semimetal, 
and $D = v_F \ell/3$ is the diffusion constant. 
The diagonal conductivity may be obtained from Eq.~\eqref{eq:29} using charge conservation
\beq
\label{eq:30}
\sigma_{xx}(\bq, \Omega) = \frac{- i \Omega e^2}{\bq^2} \chi_{00}(\bq, \Omega) = e^2 g D \frac{-i \Omega}{D \bq ^2 - i \Omega},
\eeq
which gives the standard result for the static Drude DC conductivity
\beq
\label{eq:31}
\sigma_{xx} = \lim_{\Omega \ra 0} \lim_{\bq \ra 0} \sigma_{xx}(\bq, \Omega) = e^2 g D. 
\eeq

We may also find the Hall conductivity by calculating $\chi_{xy}(\bq, \Omega)$. 
In this case, to get a well-defined result, it is important to use a regularized Hamiltonian Eq.~\eqref{eq:5}, which explicitly contains a pair of Weyl nodes of 
opposite chirality. 
The energy eigenvalues and eigenvectors of Eqs.~\eqref{eq:8}, \eqref{eq:9} are replaced by
\beq
\label{eq:32}
\epsilon_s(\bk) = s \epsilon_{\bk} = s \sqrt{v_F^2 (k_x^2 + k_y^2) + m^2(k_z)},
\eeq
and 
\beq
\label{eq:33}
| z^s_\bk \rangle = \frac{1}{\sqrt{2}} \left(\sqrt{1 + \frac{s m(k_z)}{\epsilon_\bk}}, s e^{i \varphi} \sqrt{1- \frac{s m(k_z)}{\epsilon_\bk}} \right)^T.
\eeq
Correspondingly, the SCBA Green's functions are also modified as
\beqa
\label{eq:34}
G^{R (A)}_s(\bk, \omega)&=&\frac{1}{2} \left(\frac{1}{\omega - \epsilon_\bk \pm \frac{i}{2 \tau}} + \frac{1}{\omega + \epsilon_\bk \pm \frac{i}{2 \tau}} \right), \nonumber \\
G^{R (A)}_{t x}(\bk, \omega)&=&\frac{v_F k_x}{2 \epsilon_{\bk}} \left(\frac{1}{\omega - \epsilon_\bk \pm \frac{i}{2 \tau}} - \frac{1}{\omega + \epsilon_\bk \pm \frac{i}{2 \tau}} \right), \nonumber \\
G^{R (A)}_{t y}(\bk, \omega)&=&\frac{v_F k_y}{2 \epsilon_{\bk}} \left(\frac{1}{\omega - \epsilon_\bk \pm \frac{i}{2 \tau}} - \frac{1}{\omega + \epsilon_\bk \pm \frac{i}{2 \tau}} \right), \nonumber \\
G^{R (A)}_{t z}(\bk, \omega)&=&\frac{m(k_z)}{2 \epsilon_{\bk}} \left(\frac{1}{\omega - \epsilon_\bk \pm \frac{i}{2 \tau}} - \frac{1}{\omega + \epsilon_\bk \pm \frac{i}{2 \tau}} \right). \nonumber \\
\eeqa
The static Hall conductivity is related to $\chi_{xy}$ as
\beq
\label{eq:35}
\sigma_{xy} = \lim_{\Omega \ra 0} \lim_{\bq \ra 0} \frac{e^2 v_F^2}{i \Omega} \chi_{xy}(\bq, \Omega). 
\eeq
Unlike the diagonal conductivity, which is purely dissipative and is determined entirely by impurity-scattering processes at the Fermi energy, $\sigma_{xy}$ 
in general contains both a Fermi surface and a nondissipative equilibrium contribution $\sigma_{xy} = \sigma_{xy}^I + \sigma_{xy}^{II}$.~\cite{Streda82}

We obtain
\beqa
\label{eq:36}
\sigma_{xy}^I&=& \frac{e^2 v_F^2}{\pi} \int \frac{d^3 k}{(2 \pi)^3} \textrm{Im}[G^R_s(\bk, 0) G^A_{t z}(\bk, 0)] \nonumber \\
&=&\frac{e^2 v_F^2}{2 \pi \tau} \int \frac{d^3 k}{(2 \pi)^3} \frac{m(k_z)}{\left(\epsilon_\bk^2 + \frac{1}{4 \tau^2}\right)^2}, 
\eeqa
where we have ignored diffusion vertex corrections since they do not change the result qualitatively. 
This vanishes when $1/\tau$ is much less than the bandwidth, since $m(k_z)$ changes sign at the Weyl node locations and averages to zero in the vicinity 
of each Weyl node. Thus in this limit the Fermi energy contribution to the Hall conductivity vanishes, just as in a clean Weyl semimetal.~\cite{Burkov14-2}
This may be viewed as the result of an emergent low-energy symmetry of a Weyl semimetal, namely separate conservation of the left- and right-handed charge, 
which manifests as an effective ``time-reversal symmetry", leading to a vanishing contribution to the Hall conductivity. 
While this symmetry is explicitly violated by internode scattering, it is preserved on average, which is enough to guarantee that impurity-averaged Fermi-energy 
contribution to the Hall conductivity vanishes. 

In contrast, the equilibrium part of the Hall conductivity $\sigma_{xy}^{II}$ is given by
\beqa
\label{eq:37}
&&\sigma^{II}_{xy} = \frac{e^2 v_F^2}{\pi} \lim_{\Omega \ra 0} \frac{1}{\Omega} \int_{-\infty}^{\infty} d \omega \, n_F(\omega) \int \frac{d^3 k}{(2 \pi)^3} \nonumber \\
&\times& \textrm{Im} \left[G^R_s(\bk, \omega + \Omega) G^R_{t z}(\bk, \omega) - G^R_{t z}(\bk, \omega + \Omega) G^R_s(\bk, \omega)\right]. \nonumber \\
\eeqa
This produces a result that is independent of disorder
\beq
\label{eq:38}
\sigma_{xy}^{II} = \frac{e^2}{8 \pi^2} \int_{-\pi}^{\pi} d k_z \,\textrm{sign}[m(k_z)]. 
\eeq
This needs to be supplemented by the contribution of high-energy states, which may be modeled by the same Hamiltonian as in Eq.~\eqref{eq:5}, but 
with the ``Dirac mass" that always has the same sign, for example positive. 
This finally gives
\beq
\label{eq:39}
\sigma_{xy}^{II} = \frac{e^2}{2 \pi} \frac{\cK}{2 \pi}, 
\eeq
 which is identical to the clean undoped Weyl semimetal result. 
 We thus arrive at the conclusion that even at strong disorder, but not so strong that $1/\tau$ becomes comparable to the bandwidth, 
 the Hall conductivity of a Weyl semimetal is the same as in the clean case. It also does not receive any contribution from states at the Fermi energy 
 and remains an equilibrium property, determined by all the filled states. This conclusion is consistent with previous work.~\cite{Burkov14-2,Altland16}
 \section{NLSM description of a disordered Weyl semimetal and lack of localization}
 \label{sec:3}
At this point we want to go beyond the SCBA and noninteracting diffusion modes approximation of the previous section and discuss 
the question of localization. This is most conveniently done using the matrix NLSM formalism.   
The passage to matrix NLSM is, in a sense, a bosonization transformation, which trades the fermion fields for a Hermitian matrix field $Q$. 
This is made possible by the fact that, after averaging over disorder realizations, the dynamics of the fermions becomes simpler than that of free fermions: 
instead of the continuum of particle-hole excitations of a clean Fermi gas, the low-energy excitations are sharp diffusion modes or diffusons. 
The dynamics of these modes is described by the matrix field $Q$. There exist several, largely equivalent, approaches to the matrix NLSM, we will adopt the 
Keldysh approach here,~\cite{Kamenev_book,Kamenev_Levchenko} since it is slightly more convenient in the present context (it is more straightforward to 
include coupling to the external electromagnetic field within this approach, which will be important in our discussion; Keldysh NLSM is also not limited to the Fermi 
energy, which is usually the case in standard replica NLSM formulations). 

As found in the previous section, the electromagnetic response of the Weyl semimetal contains two contributions: electric charge diffusion, 
described by Eq.~\eqref{eq:29}, and quantum anomalous Hall effect, Eq.~\eqref{eq:39} (we ignore negative magnetoresistance phenomena 
here,~\cite{Spivak12,Burkov_lmr_prb,Altland16,Burkov18-2}
which arise in an external magnetic field).
It is important that the anomalous Hall conductivity Eq.~\eqref{eq:39} is not only purely intrinsic, but is a thermodynamic equilibrium property, which, in 
a finite sample with boundaries, is associated entirely with the chiral Fermi arc surface states. This means, in particular, that it is insensitive to disorder, as 
already seen in the previous section. 

The diffusive part of the response is described by the following real-time action in the Keldysh formalism~\cite{Kamenev_book,Kamenev_Levchenko}
\beq
\label{eq:40}
i S[Q] = - \frac{\pi g}{4} \int \textrm{Tr} [D (\boldsymbol \nabla Q)^2 - 4 \partial_t Q]. 
\eeq
The field $Q$ is a $2 N \times 2 N$ Hermitian matrix, where $N$ is the number of discretized time steps in the Trotter representation of the path integral 
for the Keldysh generating function and the extra factors of $2$ correspond to the forward or backward parts of the Keldysh contour. 
The trace in Eq.~\eqref{eq:40} is thus over both the discretized time (or energy after Fourier transform) and the Keldysh contour branch indices. 
For details of the derivation of Eq.~\eqref{eq:40}, which are standard, we refer the reader to classic Refs.~\onlinecite{Kamenev_book,Kamenev_Levchenko}.

The equilibrium Hall part of the response, Eq.~\eqref{eq:39}, is contained in an extra, topological term in the matrix NLSM. 
It has been discussed before in the Weyl context in Refs.~\onlinecite{Wang15,Altland15,Altland16}, and may in principle be obtained by a gradient expansion 
of the Keldysh action.
Here we would like to derive this term by making a connection with our own earlier work on unquantized anomalies
and generalized Luttinger's theorems in metals.~\cite{Wang21} 

Let us first recall the connection between the equilibrium Hall response of Eq.~\eqref{eq:39} and the unquantized anomaly. The idea is that at low energies a Weyl semimetal with a pair of nodes has an enlarged emergent symmetry $U(1)_L \times U(1)_R$, 
corresponding to a separate conservation of left- and right-handed fermions. 
Such a symmetry, taken as a microscopic symmetry, is anomalous, and may only be realized on the surface of a 4D quantum spin Hall 
insulator, characterized by the response
\beq
\label{eq:41}
\cL = \frac{i}{6 (2 \pi)^2} (A_R \wedge d A_R \wedge d A_R - A_L \wedge d A_L \wedge d A_L), 
\eeq
which is simply two time-reversed copies of a 4D Chern-Simons term. 
Taking the Weyl nodes to be at momenta $k_z = \pm \cK$, the gauge fields $A_{R, L}$ may be decomposed as
\beq
\label{eq:42}
A_{R, L} = A \pm \pi \lambda e^z, 
\eeq
where 
\beq
\label{eq:43}
\lambda = \frac{2 \cK}{2\pi/a} = \frac{\cK a}{\pi}, 
\eeq
and 
\beq
\label{eq:44}
e^z = \frac{1}{2 \pi} d \theta^z, 
\eeq
is a translation gauge field.~\cite{Wang21} 
It is defined as the gradient of a phase $\theta^z(\br, t)$, corresponding to the family of crystal planes, perpendicular to the $z$-direction, through the relation~\cite{Volovik19,Else_QC}
\beq
\label{eq:45}
\theta^z(\br, t) = 2 \pi n, 
\eeq
where $n \in \mathbb{Z}$. 

Plugging Eq.~\eqref{eq:42} into Eq.~\eqref{eq:41} and integrating in the presence of a 3D boundary, which represents the physical Weyl semimetal, 
we obtain
\beq
\label{eq:46}
\cL = \frac{i \lambda}{4 \pi} e^z \wedge A \wedge d A, 
\eeq
which expresses the Hall conductance of $\lambda/2 \pi$ per atomic plane, i.e. precisely Eq.~\eqref{eq:39}. 

We now want to extend the above reasoning to the NLSM. 
We start by noting that the physical meaning of the matrix fields $Q$ is fluctuating density matrix
\beq
\label{eq:47}
Q^{a b}_{t t'}(\br) \sim \bar \Psi^a(\br, t) \Psi^b(\br, t'), 
\eeq
where $\Psi$ are the Grassmann fields, corresponding to electron creation and annihilation operators, $a, b$ are the Keldysh contour indices and 
summation over spin indices is implicit on the right hand side as the total electric charge is the only conserved quantity giving rise to gapless diffusion modes. 
It is important to emphasize here that in general $Q$ has an additional $2\times 2$ matrix structure, corresponding to the pseudospin indices in the Weyl Hamiltonian 
Eq.~\eqref{eq:5}. We may then expand $Q$ as 
\beq
\label{eq:47.1} 
Q = \tau^{\mu} Q_{\mu},
\eeq
where $\mu = 0, x, y ,z$. Only the spin-singlet component $Q_0$ is included in our matrix NLSM, since it corresponds to the conserved electric charge density. 
However, if the Fermi-energy part of the Hall conductivity $\sigma_{xy}^I$ was nonzero, we would need to include $Q_{x,y,z}$ into the NLSM as well, in order 
to properly capture, for example, $\chi_{x y}(\bq, \Omega)$ part of the generalized density response function. In contrast, $\sigma_{xy}^{II}$ is described entirely 
by a topological term in the NLSM, derived below, which only involves $Q_0$. 

It follows from Eq.~\eqref{eq:47} that the external electromagnetic field couples to $Q$ as
\beq
\label{eq:48}
\partial_{\mu} Q \equiv \nabla_{\mu} Q + i [\hat A_{\mu}, Q], 
\eeq
where $\mu$ are spatial coordinate indices and the matrix $\hat A_{\mu}$ is given by
\beqa
\label{eq:49}
\hat A_{\mu} = \left(
\begin{array}{cc}
A^c_{\mu} & A^q_{\mu} \\
A^q_{\mu} & A^c_{\mu}
\end{array}
\right). 
\eeqa
$A^{c, q}_{\mu}$ here are the ``classical" and ``quantum" components of the external electromagnetic gauge potentials, which correspond to symmetric 
and antisymmetric linear combinations of $A_{\mu}$ on the two branches of the Keldysh contour. 

The NLSM of a 4D quantum Hall insulator admits a topological term~\cite{Wang15}
\beq
\label{eq:50}
i S^{4D}_{top}[Q] = \frac{i}{128 \pi} \int \textrm{Tr}(Q  d Q \wedge dQ \wedge dQ \wedge dQ), 
\eeq
which is the direct analog of the Pruisken's $\theta$-term in the field theory of the 2D quantum Hall effect.
Coupling $Q$ to the chiral gauge field $A_{R,L} = \pm \pi \lambda e^z$, subtracting two time-reversed copies of the 
4D $\theta$-term, in analogy to Eq.~\eqref{eq:41}, and using $d(Q^2) = d Q Q + Q d Q = 0$ (since $Q^2 = 1$), we obtain
\beq
\label{eq:51}
i S^{4D}_{top}[Q] = \frac{\lambda}{8} \int \textrm{Tr}(e^z \wedge dQ \wedge dQ \wedge dQ). 
\eeq
Integrating this in the presence of a 3D boundary, we finally obtain
\beq
\label{eq:52}
i S_{top}[Q] = \frac{\lambda}{8} \int \textrm{Tr}(Q e^z \wedge dQ \wedge dQ), 
\eeq
which by construction is the topological term of the NLSM for a disordered Weyl semimetal.
Note that, differentiating this with respect to the translation gauge field $e^z$, gives the standard 2D Pruisken's $\theta$-term, 
with a noninteger coefficient $\lambda$, corresponding to a Hall conductance of $\lambda e^2/h$ per atomic plane. 

It is now straightforward to argue that the topological term Eq.~\eqref{eq:52} with a non-integer coefficient $\lambda$ is incompatible with localization. 
First we note that the matrix fields $Q$, representing gapless diffusion modes in a disordered Weyl semimetal, may be parametrized 
as~\cite{Kamenev_book,Kamenev_Levchenko}
\beq
\label{eq:53}
Q = T \Lambda T^{-1}, 
\eeq
where 
\beqa
\label{eq:54}
\Lambda = \left(
\begin{array}{cc}
1 & 2 F_{\omega} \\
0 & -1 
\end{array}
\right), 
\eeqa
with $F_{\omega} = 1 - 2 n_F(\omega)$, is the saddle-point value of $Q$, and $T$ are arbitrary $2N \times 2N$ matrices. 
Then, using Eq.~\eqref{eq:53}, it is easy to see that the topological term in Eq.~\eqref{eq:52} is a total derivative (assuming $d e^z = 0$, which is always true everywhere
except on dislocation lines)
\beq
\label{eq:55}
i S_{top}[T] = - \frac{\lambda}{2} \int d^3 r\, \epsilon_{\mu \nu \rho} e^z_{\mu} \textrm{Tr} [\partial_{\nu} (T \Lambda \partial_{\rho} T^{-1})]. 
\eeq

Let us now take a Weyl semimetal sample in the form of a cylinder, with the main axis along $z$ and size $L_z$. 
Suppose the sample is localized with localization length $\xi \ll L_z$ and let us consider a subsystem of our large sample 
of size $\tilde L_z$, such that $\xi \ll \tilde L_z \ll L_z$, i.e. the subsystem is still a localized insulator.
Note that it is always possible to choose such a subsystem, as long as $\xi \ll L_z$. 
The topological term for the subsystem may be written as a surface term
\beq
\label{eq:56}
i \tilde S_{top}[T] = - \frac{\lambda}{2} \int d^2 r \, \epsilon_{\mu \nu} e^z_{\mu} \textrm{Tr} \, (T \Lambda \partial_{\nu} T^{-1}). 
\eeq
This term is the action of the chiral (notice a single spatial derivative) Fermi arc on the surface of the disordered Weyl semimetal. 
We now note that the parametrization of the Goldstone mode manifold Eq.~\eqref{eq:53} is invariant under a gauge transformation
\beq
\label{eq:57}
T(\br) \ra T(\br) h(\br), 
\eeq
where $h(\br)$ is an arbitrary $2 N \times 2 N$ matrix, which commutes with $\Lambda$. 
Let $\phi$ be the azimuthal angular coordinate on the cylindrical side-surface of our Weyl semimetal sample. 
We pick 
\beq
\label{eq:58}
h(\br) = e^{i \phi \Lambda}, 
\eeq
and plug this into Eq.~\eqref{eq:56}. We obtain
\beqa
\label{eq:59}
i \tilde S_{top}[T h]&=&- \frac{\lambda \tilde N_z}{2} \int_0^{2 \pi} d \phi \, \textrm{Tr} \, (T h \Lambda \partial_{\phi} (h^{-1} T^{-1}) \nonumber \\
&=&i \tilde S_{top}[T] - \frac{\lambda \tilde N_z N }{2} \int_0^{2 \pi} d \phi \, \textrm{Tr}\, (\Lambda h \partial_{\phi} h^{-1}) \nonumber \\
&=&i \tilde S_{top}[T] + 2 \pi i \lambda \tilde N_z  N, 
\eeqa
where $\tilde N_z = \tilde L_z/a$. 
Thus for a general non-integer $\lambda$, the topological surface action is not invariant with respect to the gauge transformation of Eq.~\eqref{eq:57}. 
This implies that the representation of the topological term in the NLSM as a pure surface term is inconsistent. 
Mathematically, the problem arises due to a singularity of the parametrization in Eq.~\eqref{eq:58} on the axis of the cylinder. 
The singularity may be avoided by removing the central axis, thus creating an infinitesimally thin hole, piercing the cylinder. 
The inner surface will then carry a Fermi arc of opposite chirality, cancelling the gauge invariance violating term in Eq.~\eqref{eq:59}, if an identical gauge transformation is 
performed on both surfaces. However, if the bulk is localized, the two surfaces must be independent of each other and thus invariant under separate gauge transformations.

The physical meaning of this failure of gauge invariance is as follows. A noninteger $\lambda$ means that, just as in a clean Weyl semimetal, there must be a chiral 
Fermi arc state on the surface, with the Keldysh action given by Eq.~\eqref{eq:56}. This 2D metal may not be localized due to its chiral nature.~\cite{Balents96}
Since $\lambda$ is not an integer, this surface state may not span the whole BZ in the $z$-direction and must merge with the bulk states. 
This, in turn, implies the existence of delocalized bulk states, originating from the surface Fermi arcs. 

The internal self-consistency of the above picture requires that $\lambda$ is not renormalized by fluctuations about all the topologically-distinct classical 
saddle points of the Keldysh action and thus determines the true physical Hall conductance, not just Drude approximation to it. 
This issue has been discussed in previous work.~\cite{Wang97,Altland16}
The non-renormalization may be understood by a simple generalization of Pruisken's argument in the 2D quantum Hall effect problem,~\cite{Pruisken90,Altland10}
assuming we may ignore singular ``hedgehog" configurations of the $Q$ fields, at which the topological winding number in Eq.~\eqref{eq:52} changes abruptly at some 
coordinate $z$. 
Indeed, the Keldysh generating function may in this case be written as a sum over topological sectors 
\beq
\label{eq:60}
Z = \sum_{n = - \infty}^{\infty}e^{2 \pi i n \lambda N_z} Z_n,
\eeq
where the topological term contributes a phase factor for each winding number $n$ and $N_z = L_z/a$ is the total number of unit cells in the $z$-direction in a system of linear size $L_z$. 
Then, oddness of the Hall conductivity with respect to time reversal implies that $\sigma_{xy}$, which can be obtained from Eq.~\eqref{eq:60} by differentiating with respect to the corresponding probe gauge field components, may be written as the following ``instanton series"~\cite{Pruisken90,Altland10}
\beq
\label{eq:61}
\sigma_{xy} = \sigma_{xy}^0 + \sum_{n = 1}^{\infty} \sin(2 \pi n \lambda N_z) f_n(\sigma_{xx}^0), 
\eeq
where $\sigma_{xx}^0$ and $\sigma_{xy}^0$ are Drude values if the diagonal and Hall conductivities and $f_n$ are expansion coefficients.
Since the allowed values of $\lambda$ in a system of size $N_z$ are quantized as $\lambda = \cK a/ \pi = 2 m/N_z$, where $m \in [0, N_z/2]$ is an integer, it is clear 
that all corrections to $\sigma_{xy}^0$ vanish for all the allowed values of $\lambda$. 

This means that, in the limit $N_z \ra \infty$ the interval $\lambda \in [0,1]$ 
becomes a continuous line of fixed points of the renormalization group flow for $\sigma_{xy}$ and $\sigma_{xx}$, with $\sigma_{xx}$ not vanishing
anywhere along this line, except possibly at its end points, by the argument above. This is in contrast to the 2D quantum Hall problem, where there are stable fixed points at 
integer values of $\sigma_{xy}$ (in units of $e^2/h$), corresponding to localized quantum Hall plateaus, and unstable fixed points at half-integer values, corresponding 
to delocalized plateau transitions. This conclusion is consistent with earlier work,~\cite{Wang97,Altland16}
and with the already discussed fact that the topological term of Eq.~\eqref{eq:52} corresponds to thermal equilibrium contribution to the Hall conductivity. 
This contribution arises from all the filled states and can not be affected by fluctuations of the diffusion modes, which exist in a narrow energy window of width $1/\tau$ near the Fermi energy. It is not immediately clear how to modify this argument to include possible ``hedgehog" configurations of the matrix fields (which may be expected to be suppressed in the high conductivity, or weak coupling, limit) and thus it can not be viewed 
as a rigorous proof of non-renormalization, but only as a suggestive argument. This caveat applies to the previous work as well.~\cite{Wang97,Altland16} 
\section{Decorated domain wall construction}
\label{sec:4}
An alternative, and perhaps more physically transparent, picture of the strong disorder regime, is provided by the decorated domain wall construction.~\cite{Fu12,Ma22,Ma23}
Again, consider the simplest magnetic Weyl semimetal with a pair of nodes on the $z$-axis in momentum space at $k_z = \pm \cK$. 
As mentioned above, a single Weyl node is anomalous in the sense that it may not be on its own realized on a 3D lattice. 
It may, however, exist on the surface of a 4D quantum Hall insulator, whose electromagnetic response is described by the Chern-Simons theory 
\beq
\label{eq:1}
\cL = \frac{i}{6 (2 \pi)^2} A \wedge d A \wedge d A.
\eeq
This implies that a single Weyl node may not be localized even by strong disorder (however, note that this argument implicitly assumes that disorder is not so strong as 
to affect the topology of the 4D bulk; this is another way to state the condition already discussed above: the disorder is assumed to be not too strong, namely $1/\tau$ is less than the bandwidth). 

Returning to the physical 3D Weyl semimetal with a pair of opposite-chirality nodes, the above argument implies that localization could only arise from inter-node scattering. 
Therefore, let us focus on the inter-node part of the disorder potential and model it quasiclassically. 
Specifically, we will focus on the Fourier component of the disorder potential at the wavevector $2 \cK \hat z$, which scatters electrons directly between the two nodes 
and thus eliminates them, opening a gap. 
Let us take the potential to have a constant magnitude, but a random phase $\theta = 2 \cK {\bf R} \cdot \hat z$, where ${\bf R}$ are the Bravais lattice vectors. 
We assume that $e^{i \theta}$ averages to zero on long length scales. 
\begin{figure}
\centering
\includegraphics[width=\linewidth]{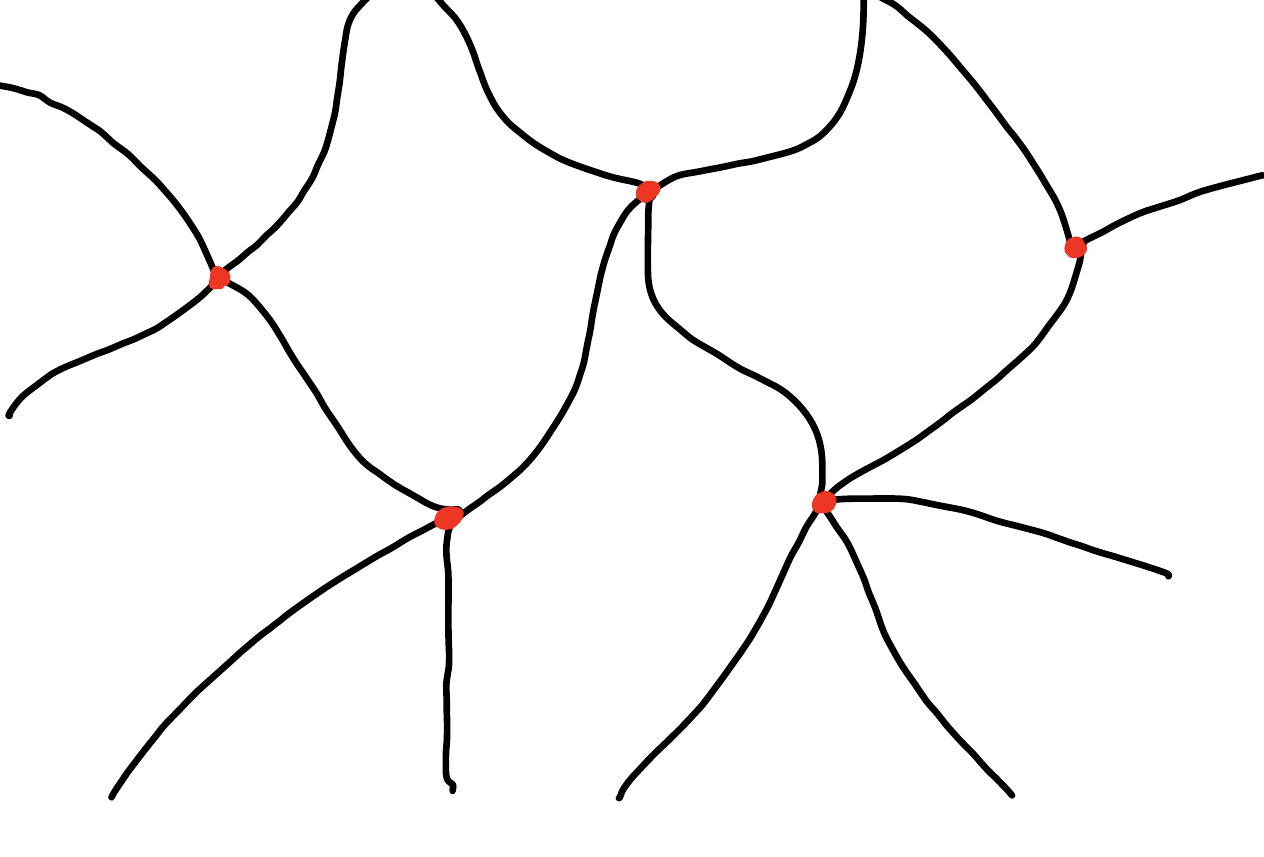}
\caption{(Color online) Quasiclassical cartoon of a disordered Weyl semimetal as a collection of domains of different values of the phase $\theta$, separated by a network 
of domain walls. Domain walls intersect at nodes, highlighted in red, which may be viewed as vortex lines, with the phase $\theta$ winding by $\pm 2 \pi$ around any path, 
encircling a vortex line.}
\label{fig:1}
\end{figure}

Quasiclassically, we may view a disordered Weyl semimetal as a collection of domains, corresponding to different values of the phase $\theta$, separated by a network 
of intersecting domain walls, as shown in Fig.~\ref{fig:1}.
Of special interest among the domain wall intersections are 1D ``vortex" lines, such that the phase $\theta$ winds by $2 \pi$ along any loop, enclosing the line. 
Such lines are significant because they trap 1D chiral modes. 
This may be seen most easily by considering the topological electromagnetic response of the disordered Weyl semimetal. 
Assuming the Hall conductivity remains unaffected by disorder, as consistent with the analysis of the previous sections, the response may be written as
\beq
\label{eq:2}
\cL = \frac{i}{8 \pi^2} d \theta \wedge A \wedge d A. 
\eeq
Indeed, taking $d \theta = 2 \cK \hat z$ gives the Hall conductivity $\sigma_{xy} = 2 \cK/ 4 \pi^2$. 

Now consider a vortex line in $\theta$, running along the $x$-direction, such that on the line $d d \theta = \pm 2 \pi$. 
In the presence of such a vortex line, Eq.~\eqref{eq:2} is not gauge invariant under $A \ra A + d \Lambda$ 
\beq
\label{eq:3}
\cL \ra \cL \pm \frac{i}{2 \pi} \Lambda (\partial_{\tau} A_x - \partial_x A_0). 
\eeq
This anomaly is cancelled by a 1D chiral mode, bound to the vortex line.~\cite{CallanHarvey}
If $e^{i \theta}$ averages to zero, restoring the broken translational symmetry of an individual domain, such vortex lines with gapless chiral modes must percolate through the entire system, which precludes localization. 

This argument needs to be modified in the case when $2 \cK = G/2$, where $G$ is a primitive reciprocal lattice vector.~\cite{Sehayek20}
In this case $\theta$ only takes two distinct values, $\theta = 0, \pi$. 
Then the response of a 2D domain wall between the domains with $\theta = 0, \pi$ is 
\beq
\label{eq:4}
\cL = \frac{i}{8 \pi} A \wedge d A, 
\eeq
which corresponds to a single massless 2D Dirac fermion (half-quantized Hall conductivity $\sigma_{xy} = 1/4 \pi$). 
When $e^{i \theta}$ averages to zero, these 2D massless Dirac states percolate, again preventing localization. 

\section{Discussion and conclusions}
\label{sec:5}
The lack of localization in a magnetic Weyl semimetal, that we have established in this paper, has a very similar origin to delocalization at a plateau transition in 2D quantum Hall systems.~\cite{Pruisken84}
The analogy becomes particularly clear if we recall that a magnetic Weyl semimetal may be viewed as an intermediate phase between a 3D integer quantum Hall insulator (i.e. a stack 
of 2D quantum Hall insulators with the Hall conductance of $e^2/h$ per layer), and an ordinary insulator. 
It is, in a sense, a broadened plateau transition, which does not have to be sharp in 3D since the Hall conductivity is not dimensionless in units of $e^2/h$, but contains a 
wavevector, which may go to zero smoothly.

The main result of our paper may then be formulated as follows. Suppose we start with a 3D integer quantum Hall insulator and add impurities. 
At strong enough disorder the 3D quantum Hall insulator phase will always be destroyed and only a trivial Anderson insulator phase with zero Hall conductivity 
will exist. Suppose disorder is not this strong and we may still distinguish two insulating phases: a 3D integer quantum Hall insulator with a Hall conductance 
of $e^2/h$ per atomic plane and a 3D ordinary insulator with zero Hall conductance (this is equivalent to the condition that $1/\tau$ is smaller than the bandwidth, mentioned before). Then we claim that there always exists an intermediate metallic (i.e. delocalized) 
phase between the two insulators, with a Hall conductance of $\lambda e^2/h$ per atomic plane, where $\lambda$ is continuously tunable between $0$ and $1$. 
The Weyl nodes themselves are always destroyed by strong enough disorder, but metallicity remains. 

We note here that there also exists a close analogy between the lack of localization in Weyl semimetals and in the surface states of weak topological insulators.~\cite{Stern12,Mong12,Fu12} In both cases gaplessness is protected by a crystal symmetry (translation in the Weyl case), which is violated by disorder, 
but is restored upon averaging. Average symmetry is then sufficient to protect against localization.

An interesting question is whether this statement may be generalized to other types of topological semimetals. 
We believe that the answer is yes, at least in some cases. 
While it should be possible to generalize the NLSM-based argument, it is much easier to use the decorated domain wall construction of Section~\ref{sec:4}.  
For example, let us take the simplest nonmagnetic Weyl semimetal with broken inversion symmetry, which contains two pairs of nodes, related by time-reversal. 
In this case, if we take the disorder potential to be time-reversal symmetric, it will only couple nodes within each pair. Then one may use the argument of
Section~\ref{sec:4} separately for each pair of nodes and arrive at the conclusion that localization is always avoided, as long as the disorder is not strong enough 
to destroy the ``parent" 3D weak spin Hall insulator state, which is realized when the nodes in each pair are annihilated at the edges of the BZ. 

The unquantized anomaly viewpoint, discussed in this paper, and in Refs.~\onlinecite{Gioia21,Wang21,Sau24,Hughes24} in a lot more detail, connects
topological semimetals with ordinary metals, demonstrating that topological responses of the semimetals, such as the fractional Hall conductance per 
atomic plane, may be viewed as generalizations of the Luttinger's theorem in ordinary metals. On the other hand, the lack of localization even in strong 
disorder, differentiates topological semimetals from ordinary metals, which are not protected against localization, despite topological origin of the 
Luttinger's theorem. 

Experimentally, the lack of localization in 3D topological semimetals is not as easily observable as in the case of the quantum Hall plateau transition in 2D. 
In 2D all states are localized even in arbitrarily weak disorder potential, except at the plateau transition (if we ignore electron-electron interactions). In contrast, in 3D localization only occurs at strong enough disorder, with a non-universal critical strength, which depends on the details of a given material. In particular, it may be comparable to the disorder strength, necessary to destroy the 3D integer quantum Hall insulator.
Our results are thus of a more conceptual than practical value. However, one can say that topological semimetals may generally be expected to be more conductive than ordinary 3D metals with comparable density of states and impurity density, due to the lack of localization tendency, demonstrated in this paper. 

\begin{acknowledgments}
We acknowledge helpful discussions with Eslam Khalaf, Michael Smith and Chong Wang. Financial support was provided by the Natural Sciences and Engineering Research Council (NSERC) of Canada.
AAB was also supported by Center for Advancement of Topological Semimetals, an Energy Frontier Research Center funded by the U.S. Department of Energy Office of Science, Office of Basic Energy Sciences, through the Ames Laboratory under contract DE-AC02-07CH11358. 
Research at Perimeter Institute is supported in part by the Government of Canada through the Department of Innovation, Science and Economic Development and by the Province of Ontario through the Ministry of Economic Development, Job Creation and Trade.
This research was also supported in part by grant NSF PHY-2309135 to the Kavli Institute for Theoretical Physics (KITP). 
\end{acknowledgments}
\bibliography{references}
\end{document}